\RequirePackage{fix-cm}
\documentclass[smallcondensed]{svjour3}     % onecolumn (ditto)
\smartqed  % flush right qed marks, e.g. at end of proof
\usepackage{graphicx}
\begin{document}

\title{WHY IS GEV PHYSICS RELEVANT IN THE AGE OF THE LHC?
\thanks{The work was authored in part by Jefferson Science Associates, LLC under U.S. DOE Contract No. DE-AC05-06OR23177.}
%Grants or other notes
%about the article that should go on the front page should be
%placed here. General acknowledgments should be placed at the end of the article
}

%\subtitle{Do you have a subtitle?\\ If so, write it here}

\titlerunning{GeV Physics}        % if too long for running head

\author{M.R. Pennington
}

%\authorrunning{Short form of author list} % if too long for running head

\institute{ \at
         Theory Center,
Thomas Jefferson National Accelerator Facility,\\
12000 Jefferson Avenue, Newport News, VA 23606, U.S.A. \\
              Tel.: +1-757-269-6026
              \email{michaelp@jlab.org}            
}

\date{Received: date / Accepted: date}
% The correct dates will be entered by the editor

\maketitle

\begin{abstract}
The contribution that Jefferson Lab has made, with its 6 GeV electron beam, and will make, with its 12 GeV upgrade,  to our understanding of the way the fundamental interactions work, particularly strong coupling QCD, is outlined. This physics at the GeV scale is essential even in TeV collisions.
% \subclass{MSC code1 \and MSC code2 \and more}
\end{abstract}

\section{Why CEBAF?}
\label{intro}
At Jefferson Lab, we study all the interactions of the Standard Model and even those Beyond, but at a quite different energy regime than the LHC: GeV rather than TeV. Precision parity violation experiments measure the weak charge of the proton at low momentum scales, and searches go on for heavy photons. Both set limits on what is beyond the Standard Model. However, the focus of this talk is precision study of the strong interaction using electromagnetic probes.

At Jefferson Lab~\cite{whitepaper}, where our electron machine (known as CEBAF) is being upgraded to 12 GeV, we study the spectrum and  structure of hadrons, and in turn how these hadrons, particularly nucleons, build nuclei and so determine the properties of the matter of which we are made. These properties reflect the nature of the strong interaction, governed by QCD, in the strong coupling regime: a regime that is responsible for colour confinement and chiral symmetry breaking that shape the dynamics of hadrons. The QCD Lagrangian incorporates all the features required for perturbative calculations to the first few orders that can and do describe hard scattering processes, involving interactions over ranges much smaller than the size of a hadron, like the anti-proton. That is the success dramatically enabled by the asymptotic freedom of QCD.  But what happens at longer distances, like 1~{\em fm}, when the interactions of quarks and gluons become strong? Then new phenomena arise. {\em Up} and {\em down} quarks that were very light when travelling over 0.01~{\em fm} become very much heavier.
Over a fermi they become confined, building hadrons and fixing all their properties and interactions. Experiments at the LHC aim to use the very short distance interactions of quarks, antiquarks and gluons to create forms of matter not previously free since the Big Bang: the Higgs, perhaps supersymmetric partners of all the particles we know, glimpses of extra dimensions,{\em etc}. There protons collide. The hadronic debris triggering the huge detectors at the LHC are pions, kaons and nucleons: objects of the size of a fermi. Consequently, to decipher what the LHC observes we have to understand the description of quarks and gluons inside hadrons, both how they break up, but also how they get back together again. These processes even at TeV energies are governed by physics at the GeV scale. It is this we study in fine detail at Jefferson Lab.~\cite{whitepaper} 

Both the spectrum and structure of hadrons are direct reflections of strong coupling QCD. Our understanding of how these properties emerge is studied both by calculation and by experiment. In these times of stretched financial resources, it is sometimes argued that since we claim to know the Lagrangian of the strong interaction, we can learn everything by performing calculations either on the lattice or in the continuum. While considerable progress has been made, as we will touch on, for many aspects we still have some way to go to gain a quantitative understanding. Rather we will follow the alternative attack provided by experiment. Remarkably quarks know how to solve the equations of strong coupling QCD without the need of super-computers and approximations.  Consequently, experiment provides us with powerful information about how QCD really works.

Baryons (and here at LEAP, antibaryons) play a special role in the study of QCD. They are intimately tied to the non-Abelian nature of QCD. Mesons don't care how many colours there are, but for baryons, the minimum number of quarks of which they are made is, of course, identical to  the number of colours. Thus, as appropriate for a meeting dedicated to antimatter, the simplest colour singlet wavefunction of an antiproton with colours red ({\em r}), green ({\em g}) and blue ({\em b}) is:
\begin{eqnarray}
\nonumber
\large|\,{\overline {\boldmath p}}\,\rangle\; =&\large\frac{1}{\sqrt{6}}& \large|\left[\,{\overline{\boldmath{u_r\,u_g\,d_b}}}\,+\,{\overline{\boldmath{u_g\,u_b\,d_r}}}\,+\,{\overline{\boldmath{u_b\,u_r\,d_g}}}\right.\\[3mm]
&&-\,\left.\large{\overline{\boldmath{u_g\,u_r\,d_b}}}\,-\,{\overline{\boldmath{u_r\,u_b\,d_g}}}\,-\,{\overline{\boldmath{u_b\,u_g\,d_r}}}\,\right]\,\rangle
\end{eqnarray}

%%%%%%%%%%%%%%%%%%%%%%%%%%%%%%%%%%%%%%%%%%%%%%%%%%%%%%%%%%%%%%%%%%%%%%%%%%%%5

\begin{figure}[h]
\begin{center}
\includegraphics[width=0.95\textwidth]{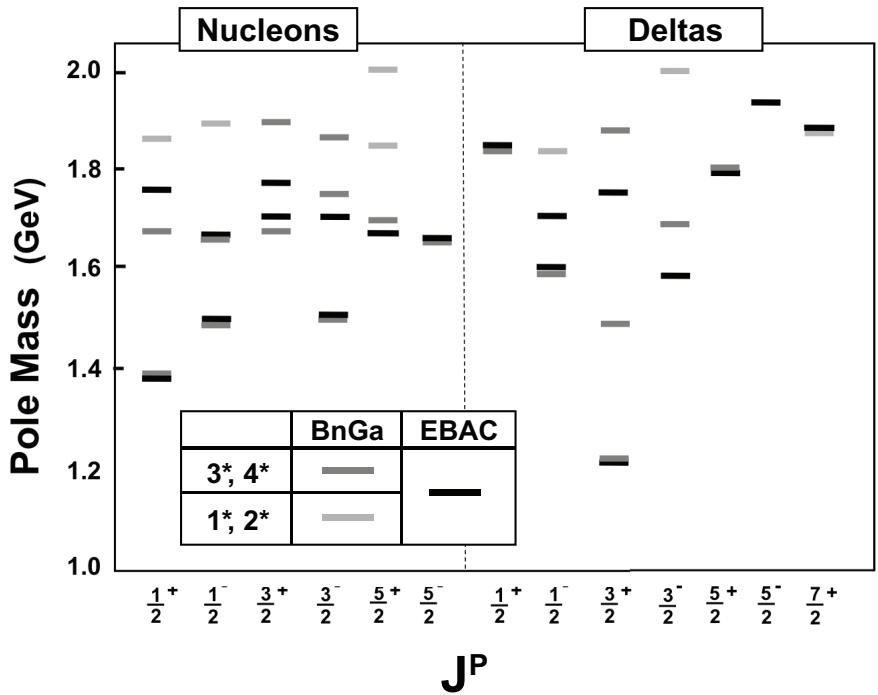}
%\vspace{-2mm}
  \caption{$N^*$ and $\Delta^*$ spectra, labeled by their spin and parity as $J^P$ along the abscissa, and the real part of the resonance pole positions along the ordinate, from the EBAC~\cite{ebac2} and Bonn-Gatchina~\cite{bn-ga} analyses. For the EBAC ({\it aka} ANL-Osaka) analysis all the states have $3^*-4^*$ provenance, while Bonn-Gatchina also include those with $1^*-2^*$ ratings, according to the legend shown.
Note the tendency of some  $N^*$'s and $\Delta^*$'s to appear in parity pairs as their mass increases above 1800 MeV.}
\end{center}
\vspace{-4.5mm}
\end{figure}
%%%%%%%%%%%%%%%%%%%%%%%%%%%%%%%%%%%%%%%%%%%%%%%%%%%%%%%%%%%%%%%%%%%%%%%%%%%%%
\noindent Even if this is really all there is to a ground state (anti-)baryon, what about their excitations? The 6~GeV program at JLab has added significantly to knowledge of the light baryon spectrum. Ten years ago, there seemed to be many  \lq\lq missing'' states. Baryons that the quark model with three quark degrees of freedom would have us expect~\cite{capstick},  and so if absent pointing perhaps to a simpler diquark-quark structure. However, we now know these states were not missing~\cite{sarantsev}, but just \lq\lq dark'' in the $\pi N$ and $\pi\pi N$ channels that were the main source of information. But now add to these $KY$ decays and a wealth of photoproduction data with polarized photons on polarized protons (and neutrons too)~\cite{bn-ga2,bn-ga3} from CBELSA@Bonn, MAMI@Mainz and Hall B@JLab and these states are starting to appear. Partial wave analyses of these data have been most comprehensively performed in two parallel treatments: one by the EBAC team of Lee {\it et al.}~\cite{ebac,ebac2} in terms of an underlying Lagrangian of hadronic interactions, and the other by the Bonn-Gatchina team~\cite{bn-ga} with a more flexible multi-channel amplitude analysis treatment. The latest results from these two groups for $N^*$'s and $\Delta^*$'s are shown in Fig.~1. 
%%%%%%%%%%%%%%%%%%%%%%%%%%%%%%%%%%%%%%%%%%%%%%%%%%%%%%%%%%%%%%%%%%%%%%%%%%%%5
\begin{figure}[th]
\begin{center}
  \includegraphics[width=.78\textwidth]{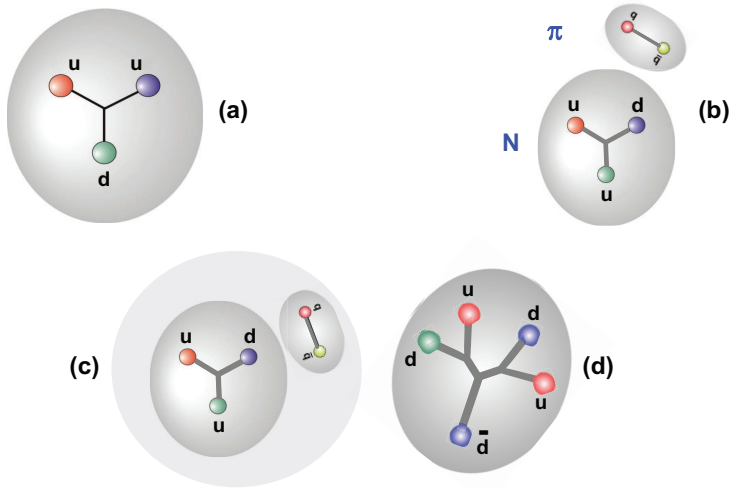}
  \caption{Cartoon of the possible Fock components (a-d) of some excited baryon, for instance the $N^*(1440)$. It almost certainly has components (a) and (b), but the relative amounts of (a-d) awaits to be determined for the Roper, or any other excited, baryon.}
\end{center}
\vspace{-5mm}
\end{figure}
%%%%%%%%%%%%%%%%%%%%%%%%%%%%%%%%%%%%%%%%%%%%%%%%%%%%%%%%%%%%%%%%%%%%%%%%%%%%%

For the most part these agree, but the Bonn-Gatchina treatment allows channels like $\pi\pi N$ to be incorporated more readily, and consequently their analysis extends to higher baryon masses. 
That seems to reveal parity doublets~\cite{bn-ga-doublet}. Is this just an accident around 1.9-2 GeV, or is it the start of something not predicted in the quark model~\cite{capstick}? We need to identify the flavoured companions of all these excited baryons. The $Y^*$'s and $\Xi^*$'s are expected to be narrower: having always to produce one (or more) kaon in their decays, for which they have much less phase space. Plans are afoot to search for these.

What we want to know for all the states in Fig.~1, is their Fock space decomposition. It is clearly more complex than that of $qqq$ of Fig.~2a and Eq.~(1).  Each state decays to $\pi N$, $\pi\pi N$, $K\Lambda$, $\ldots$, so spends some of its time in a multi-hadron configuration, like Fig.~2b. What teaches us about the way confinement dynamics works is whether this is just a meson cloud around a $qqq$-core (Fig.~2c) or something intrinsically more complicated, like Fig.~2d. The EBAC analysis (by what is now  known as the ANL-Osaka collaboration)~\cite{ebac} can express this in terms of the \lq\lq bare'' hadronic seeds of its Lagrangian. Probing excited states with virtual photons can in principle look inside each state and picture its composition. However, since these states are not stable we have instead to deduce their structure from a more fuzzy image in which the probing photon transforms the target nucleon into an exclusive $N^*$~\cite{mokeev1,mokeev2}. The study of transition formfactors is a major program to investigate their composition.

%%%%%%%%%%%%%%%%%%%%%%%%%%%%%%%%%%%%%%%%%%%%%%%%%%%%%%%%%%%%%%%%%%%%%%%%%%%%5
\begin{figure}[b]
\begin{center}
\includegraphics[width=0.95\textwidth]{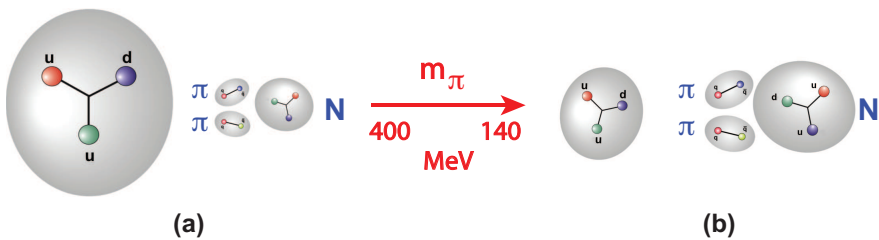}
\caption{Cartoon of how the relative amounts of $qqq$-core and $\pi\pi N$ components in an excited baryon may change as the pion mass is decreased from (a) 400 to (b) 140 MeV.} 
\end{center}
\vspace{-2.mm}
\end{figure}
%%%%%%%%%%%%%%%%%%%%%%%%%%%%%%%%%%%%%%%%%%%%%%%%%%%%%%%%%%%%%%%%%%%%%%%%%%%%%
Alternatively, we can compute the spectrum from strong QCD. 
Considerable progress has been made in continuum studies of hadronic properties, see for instance the reviews~\cite{cdr}. This approach using the system of  Schwinger-Dyson equations for quarks, gluons and their interactions, and suitable Bound State equations to couple these together, has the beautiful feature of being applicable at any quark mass. In particular, chiral symmetry breaking that is so important for the behaviour of the low energy world is naturally embodied. On the lattice, calculation is presently only possible with precision with heavy pions~\cite{edwards1}. Then the world looks very like the constituent quark model. Perhaps this is not too surprising, since with 400 or 500 MeV pions, the quark core of  the $N^*$'s and $\Delta^*$'s is only supplemented by small $\pi N$, $\pi\pi N$, {\it etc.} components, Fig.~3a. As calculations advance towards physical mass pions, these components will inevitably become larger, as sketched in Fig.~3b. It could be that even if the 3-quark core states with opposite parity (above say 2~GeV) are displaced in mass, their differing decay patterns may make the true hadrons almost degenerate, as the Bonn-Gatchina analysis~\cite{bn-ga-doublet} appears to suggest, Fig.~1. Decays are an integral part of the make-up of all hadrons, fixing their content, and shifting their masses. The EBAC results on the Roper suggests its stable core is many hundreds of MeV heavier than 1440 MeV~\cite{ebacpoles}.

 The story of the $X,Y, Z$ mesons, discussed here by Stephen Olsen~\cite{olsen}, in charmonium, bottomonium, and perhaps strangeonium is an intriguing window on how hadrons really have a multiquark composition, and their decay dynamics shapes their properties too. Detailed study of their decays has the capacity to teach us whether they are bound by interhadron forces, as for molecular states, or by interquark forces that bind tetraquark mesons.  Of course, it is through such windows we hope to understand how  confinement really works in practice.  
This is an exciting view we watch with eager anticipation as it unfolds.

The plans for JLab at 12 GeV~\cite{whitepaper} include a new experimental Hall D with the GlueX detector dedicated to searching in mutihadron final states for unambiguous evidence for resonances with quantum numbers that prove gluons must contribute to their $J^{PC}$ quantum numbers: quantum numbers not allowed by simple ${\overline q}q$ configurations. The lightest of these  \lq\lq exotic'' mesons is expected to have $1^{-+}$ quantum numbers~\cite{dudek}. Hints of such states at 1400 and 1600 MeV have been glimpsed by GAMS~\cite{gams} in its $\pi\eta$ decays,  by VES in $\pi\eta$ and $\pi\eta^\prime$~\cite{ves}, by BNL-E852 in  $\pi\eta$ and $3\pi$ decays~\cite{chung,dzierba}, and more recently at 1600~MeV by CLEO-c in the decay of the $\chi_{c1}$~\cite{cleoc} and by COMPASS at CERN~\cite{compass0,compass1}. All see an enhancement, but is this a resonance? Is there a pole in the complex energy plane? The required analytic continuations demand knowledge not only of the energy variation of the modulus, but also of the phase of the relevant partial wave. Such phase variation (relative to some understood wave, like $2^{++}$) is complicated by the role of other production mechanisms, like the Deck effect, that contribute to the same final state. These issues are the essence of studies of the nearly 100 million events in the $3\pi$ channel taken by the COMPASS experiment. These are complications that will be at least, if not more important, in polarized photoproduction with GlueX. Ways to deal with such complexities is a target project  of the new JLab Physics Analysis Center.

 If hybrid mesons exist then we need to find them not only with $J^{PC}\,=\,1^{-+}$, but with their nearby companion $0^{+-},\ 2^{+-}$ quantum numbers too, in not just one flavour combination, but in whole multiplets. It is these structures that will tell us that these really are ${\overline q}qg$ hybrid states, and not multiquark states, like ${\overline{qq}}qq$. A long sought sign of gluonic degrees of freedom are hadron states known as {\it glueballs}. Thirty years ago we thought they would be readily identifiable, the lightest having scalar quantum numbers. However, their decays to pions, kaons, etas,$\ldots$, all require coupling to quarks and hence mixing with conventional ${\overline q}q$ states with zero flavour quantum numbers. Once again it is their decays that colour and shape what is observed in experiments. Consequently, their identification, as super-numerary states, is best done in processes where many channels are studied at once. An example is the central production of di-meson final states in high energy $pp$ collisions. Again  data from COMPASS, when partial wave analysed in several channels simultaneously~\cite{compass2}, may yet prove a good hunting ground. Excited $N^*$s, in which glue contributes to their $J^P$, are also predicted. However, they have no {\it unusual} quantum numbers. Recent lattice calculations~\cite{edwards2} give these a mass $\sim 1$~GeV heavier than their largely $qqq$ companions. At present, only an excess of excited baryons would suggest their existence. However, future calculations might predict they have distinctive decay patterns, which might make searching for these a little less problematic. 
%Consequently, the focus is on hybrid and gluish mesons.

\section{Hadron structure}
We now turn to the structure of hadrons, especially nucleons. It is a mission of JLab  to determine the momentum, flavour, and angular momentum distributions of partons. This is, of course, a continuation of a long established program of deep inelastic scattering. Studies that started at SLAC, {\it e.g.}~\cite{holt}. What JLab will add is precision in the valence region with both proton and simple nuclear targets (like $^3 He$ and $^3 H$) that allow scattering on a neutron to be separated. These processes are studied with polarized photons and with polarized targets, that allow spin distribution functions to be determined. As   illustrated in Fig.~4, our present knowledge of even the  unpolarized $u$ and $d$ distributions (pdfs) is rather poor in the valence region of larger~$x$~\cite{cj11}. However, after 
12 GeV running at JLab, this should dramatically improve and take the modelling out of current flavour separation.  Of course, the very concept of parton distributions is underpinned by the asymptotically free nature of QCD: quarks and gluons interact only weakly, and so the struck parton interacts with the probing photon independently of its companions. 
%%%%%%%%%%%%%%%%%%%%%%%%%%%%%%%%%%%%%%%%%%%%%%%%%%%%%%%%%%%%%%%%%%%%%%%%%%%%5
%\vspace{-3mm}
\begin{figure}[b]
\begin{center}
\includegraphics[width=0.72\textwidth]{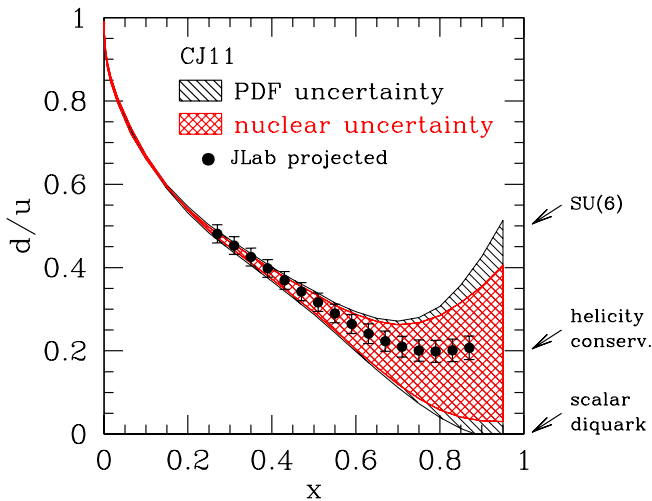}
\vspace{-1.8cm}
\caption{The ratio of $d$ to $u$ parton distributions, as a function of $x$, the longitudinal momentum fraction of the parton in the infinite momentum frame, at a photon virtuality $Q^2 \sim 10$ GeV$^2$ from the recent analysis by the CTEQ-JLab group, {\em CJ}~\cite{cj11}. The shaded bands show the current uncertainties. The data points indicate the size of the errors to be expected after the 12 GeV running at JLab~\cite{whitepaper}. The $x \to 1$ limits of the ratio in well-known models are arrowed.}
\end{center}
\vspace{-2.mm}
\end{figure}
%%%%%%%%%%%%%%%%%%%%%%%%%%%%%%%%%%%%%%%%%%%%%%%%%%%%%%%%%%%%%%%%%%%%%%%%%%%%%

%%%%%%%%%%%%%%%%%%%%%%%%%%%%%%%%%%%%%%%%%%%%%%%%%%%%%%%%%%%%%%%%%%%%%%%%%%%%5

\begin{figure}[t]
\begin{center}
\includegraphics[width=0.95\textwidth]{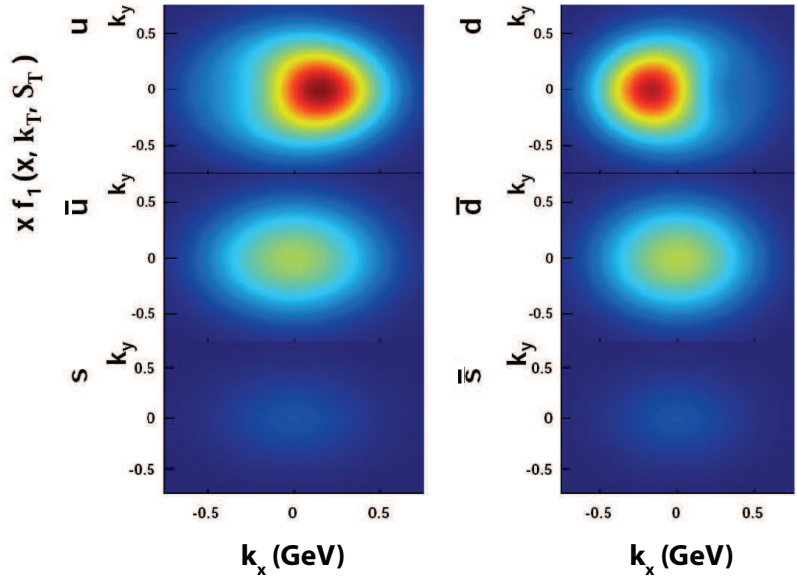}
\caption{The flavour separated $u,\ d,\ s$ (and their antiquark) transverse momentum distributions in the $k_x-k_y$ plane at a typical longitudinal ($z$-directon) momentum fraction $x \simeq 0.1$ for a proton polarized in the $y$-direction as expected from 12 GeV running at JLab. Notice how these {\em Sivers} distributions have $u$ and $d$ distributions skewed in opposite directions and perpendicular to the direction of the proton's polarization $S_T$~\cite{whitepaper}.}
\end{center}
\vspace{-2.mm}
\end{figure}
%%%%%%%%%%%%%%%%%%%%%%%%%%%%%%%%%%%%%%%%%%%%%%%%%%%%%%%%%%%%%%%%%%%%%%%%%%%%%

The question of how the spin of the proton is built from  the angular momentum of its partons naturally requires knowledge not only of the longitudinal motion of these partons, but also their transverse distributions, either in configuration or momentum space, {\it e.g.}~\cite{spin1}. Such information is encoded in Generalized Parton and Tranverse Momentum-dependent Distributions, respectively. Importantly, these 3-dimensional distributions are linked on integration to formfactors and the usual pdfs.
While GPDs can be explored in deeply virtual processes like Compton scattering or exclusive meson production, Transverse Momentum Distributions (TMDs) appear in the asymmetries of Semi-Inclusive Deep Inelastic Scattering (SIDIS). Their extraction requires knowledge of how after the quark has been struck it gets back together with others to build the specific  hadron that is observed. Thus we need fragmentation functions too~\cite{prokudin}. Quantities that come from other experiments at $e^+e^-$ colliders, for instance. 

Results from HERMES~\cite{hermes},  COMPASS~\cite{compass-tmd} and JLab~\cite{spin1,whitepaper}
have shown that non-zero TMDs can be determined. JLab with a whole suite of approved experiments~\cite{whitepaper} will map out in detail multi-dimensional SIDIS production of pions and kaons.  Of the eight lowest twist distributions, we use the Sivers function, $f_{1T}^{\perp}(x, k_T, Q)$, here for illustration. The distribution in the transverse momentum plane  of unpolarized quarks in an unpolarized nucleon is naturally symmetric. However, when the nucleon is transversely polarized, the quark distributions become skewed as illustrated in Fig.~5. The Sivers function describes unpolarized quarks in a transversely polarized nucleon. We see {\em up} quarks skew in one direction, {\em down} in the opposite. This reflects the orbital motion of the quarks. Studies in Deeply Virtual Compton Scattering and Meson Production likewise teach us that the orbital angular momentum of the $u$ and $d$ quarks in a nucleon are significant, but almost cancel when added. This continues to leave a mismatch between the spin of the nucleon ($\hbar/2$) and the total angular momentum carried by  quarks ($\sim 0.15 \hbar$), indicating that the  angular momentum in the gluon's motion should fill the gap. An intense debate is going on about how to define such a gluonic quantity in a measurable way. The current state of the debate is summarised in Ref.~\cite{spin2}. 
This activity will hopefully inform experiments to come.

\section{Prospects}
As with the investigation of the spin of the nucleon,
all studies of strong interaction phenomena at JLab aim to expose the correlations between partons, whether in the nucleon itself, in its excitations or inside nuclei.
The way each parton knows of the others reflects both confinement and the breaking of chiral symmetry. Experiment and theory march hand-in-hand.
Only when our ability and capacity to compute these consequences of QCD matches anything near the precision of the data to come from CEBAF at Jefferson Lab with the 12 GeV upgrade,  will we be able to claim that we really understand the  fermi size objects that the LHC collides and detects, and which build the visible universe. 
This is what inspires the 12 GeV adventure for the next decade or more.

\begin{acknowledgements}
It is a pleasure to thank the organisers, especially Tord Johansson,
 for the invitation to give this talk. Combining the scientific presentations with G\"osta Ekspong's wonderful memories of the discovery of the antiproton at Berkeley, and with Karin Sch\"onning and her friends tribute to ABBA, made this a most enjoyable and memorable conference.
\end{acknowledgements}


\begin{thebibliography}{}
\bibitem{whitepaper} J.~J.~Dudek, R.~Ent, R.~Essig, K.~S.~Kumar, C.~Meyer, R.~D.~McKeown, Z.~E.~Meziani and G.~A.~Miller, M.~R.~Pennington, D.~G.~Richards, L.~Weinstein, G.~Young and S.~Brown,
  Physics Opportunities with the 12 GeV Upgrade at Jefferson Lab,
  {\it Eur.\ Phys.\ J.\ A} {\bf 48}, 187 (2012)
  [arXiv:1208.1244 [hep-ex]].

\bibitem{capstick} 
  S.~Capstick and N.~Isgur,
  Baryons in a Relativized Quark Model with Chromodynamics,
 {\it Phys.\ Rev.\  D} {\bf 34}, 2809 (1986).
 
  S.~Capstick and W.~Roberts,
  Quasi two-body decays of nonstrange baryons,
  {\it Phys.\ Rev.\  D} {\bf 49}, 4570 (1994)
  [arXiv:nucl-th/9310030].
\bibitem{sarantsev}
A.~V.~Sarantsev,
Bonn-Gatchina partial wave analysis: Search for missing baryon states,
 {\it Acta Phys.\ Polon.\ Supp.}\  {\bf 3}, 891 (2010).

 V.~D.~Burkert,
  Evidence of new nucleon resonances from electromagnetic meson production,
  {\it EPJ Web Conf.}\  {\bf 37} (2012) 01017
  [arXiv:1209.2402 [nucl-ex]].  
\bibitem{bn-ga2}
A.~V.~Anisovich, R.~Beck, E.~Klempt, V.~A.~Nikonov, A.~V.~Sarantsev and U.~Thoma,
  Pion- and photo-induced transition amplitudes to $\Lambda K$, $\Sigma K$, and $N\eta$,
  {\it Eur.\ Phys.\ J.\ A} {\bf 48}, 88 (2012)
  [arXiv:1205.2255 [nucl-th]].
\bibitem{bn-ga3}
  A.~Thiel, A.~V.~Anisovich, D.~Bayadilov, B.~Bantes, R.~Beck, Y.~.Beloglazov, M.~Bichow and S.~Bose {\it et al.}, Well-established nucleon resonances revisited by double-polarization measurements,
  {\it Phys.\ Rev.\ Lett.}\  {\bf 109}, 102001 (2012)
  [arXiv:1207.2686 [nucl-ex]].


\bibitem{ebac2}
  H.~Kamano, S.~X.~Nakamura, T.~-S.~H.~Lee and T.~Sato, 
  Nucleon resonances within a dynamical coupled-channels model of $\pi N$ and $\gamma N$ reactions,
  arXiv:1305.4351 [nucl-th].



\bibitem{bn-ga} A.~V.~Anisovich, R.~Beck, E.~Klempt, V.~A.~Nikonov, A.~V.~Sarantsev and U.~Thoma,
  Properties of baryon resonances from a multichannel partial wave analysis,
  {\it Eur.\ Phys.\ J.\ A} {\bf 48}, 15 (2012)
  [arXiv:1112.4937 [hep-ph]].

\bibitem{ebac}
 H.~Kamano and T.~-S.~H.~Lee,
  EBAC-DCC Analysis of World Data of $\pi N$, $\gamma N$, and $N(e,e')$ Reactions,
  {\it AIP Conf.\ Proc.}\  {\bf 1432} 74 (2012)
  [arXiv:1108.0324 [nucl-th]].
\bibitem{bn-ga-doublet}
  A.~V.~Anisovich,  E.~Klempt, V.~A.~Nikonov, A.~V.~Sarantsev, H.~Schmieden and U.~Thoma, Evidence for a negative-parity spin-doublet of nucleon resonances at 1.88\,GeV,
  {\it Phys.\ Lett.\ B} {\bf 711}, 162 (2012)
  [arXiv:1111.6151 [nucl-ex]].


\bibitem{mokeev1}
  I.~G.~Aznauryan and V.~D.~Burkert,
  Electroexcitation of nucleon resonances,
  {\it Prog.\ Part.\ Nucl.\ Phys.}\  {\bf 67}, 1 (2012)
  [arXiv:1109.1720 [hep-ph]].

\bibitem{mokeev2} I.~G.~Aznauryan {\it et al.} [CLAS],
 Electroexcitation of nucleon resonances from CLAS data on single pion electroproduction,
 {\it  Phys.\ Rev.\ C} {\bf 80}, 055203 (2009)
  [arXiv:0909.2349 [nucl-ex]].

V.~I.~Mokeev, I.~G.~Aznauryan and V.~D.~Burkert, Nucleon Resonance Electrocouplings from the CLAS Data on Exclusive Meson Electroproduction off Protons,
  arXiv:1109.1294 [nucl-ex].
\bibitem{cdr} 
I.~C.~Cloet, C.~D.~Roberts and D.~J.~Wilson,
Baryon properties from Continuum QCD,
{\it AIP Conf.\ Proc.}\ {\bf 1388}, 121 (2011).

  G.~Eichmann,
  From quarks and gluons to baryon form factors,
  {\it Prog.\ Part.\ Nucl.\ Phys.}\  {\bf 67}, 234 (2012).


\bibitem{edwards1}  R.~G.~Edwards, J.~J.~Dudek, D.~G.~Richards and S.~J.~Wallace,
  Excited state baryon spectroscopy from lattice QCD,
{\it Phys.\ Rev.\ D} {\bf 84}, 074508 (2011).

\bibitem{ebacpoles}
H.~Kamano, S.~X.~Nakamura, T.~S.~Lee and T.~Sato [EBAC],
Extraction of $P_{11}$ resonances from $\pi N$ data,
  {\it Phys.\ Rev.\  C} {\bf 81}, 
065207 (2010).
[arXiv:1001.5083 [nucl-th]].
\bibitem{olsen} S.~L.~Olsen, these proceedings.

\bibitem{dudek} J.~J.~Dudek, R.~G.~Edwards, M.~J.~Peardon, D.~G.~Richards and C.~E.~Thomas,
  Toward the excited meson spectrum of dynamical QCD,
 {\it Phys.\ Rev.\ D} {\bf 82}, 034508
(2010)  [arXiv:1004.4930 [hep-ph]].

  J.~J.~Dudek,  R.~G.~Edwards, B.~Joo, M.~J.~Peardon, D.~G.~Richards and C.~E.~Thomas,
  Isoscalar meson spectroscopy from lattice QCD,
  {\it Phys.\ Rev.\ D} {\bf 83}, 111502 (2011)
  [arXiv:1102.4299 [hep-lat]].
\bibitem{gams} D.~Alde {\it et al.}  [IHEP-Brussels-Los Alamos-Annecy(LAPP) Collaboration],  Evidence for a $1^{-+}$ Exotic Meson,
  {\it Phys.\ Lett.\ B} {\bf 205}, 397 (1988).
\bibitem{ves} G.~M.~Beladidze {\it et al.} [VES], Study of $\pi^- N\to\eta \pi^- N$ and $\pi^- N \to\eta^\prime \pi^- N$ reactions at 37-GeV/c, {\it Phys.\ Lett.\ B} {\bf 313}, 276 (1993).

  Y.~P.~Gouz {\it et al.}  [VES. Collaboration],
  Study of the wave with $J^{PC} = 1^{-+}$ in the partial wave analysis of $\eta^\prime \pi^-$, $\eta \pi^-$, $f_1 \pi^-$ and $\rho^0 \pi^-$ systems produced in $\pi^- N$ interactions at $p (\pi^-) = 37$~GeV/c, {\it AIP Conf.\ Proc.}\  {\bf 272} 572 (1993).

%; D.~V.~Amelin {\it et al.} [VES] Proc. Hadron 2001, AIP\ Conf. \Proc. {\bf 619}, 143 (2002).
  
\bibitem{chung}D.~R.~Thompson {\it et al.} [BNL-E852],  Phys. Rev. Lett.\ {\bf 79}, 1630 (1997).

 S.~U.~Chung {\it et al.} [BNL-E852], Evidence for exotic $J^{PC} = 1^{-+}$ meson production in the reaction $\pi^- p\to \eta \pi^- p$ at 18-GeV/c, {\it Phys.\ Rev.\ D} {\bf 60}, 092001 (1999) [hep-ex/9902003].

\bibitem{dzierba} A.~R.~Dzierba, J.~Gunter, S.~Ichiriu, R.~Lindenbusch, E.~Scott, P.~Smith, M.~R.~Shepherd and S.~Teige {\it et al.}, 
  A Study of the $\eta \pi^0$ spectrum and search for a $J^{PC} = 1^{-+}$ exotic meson, {\it Phys.\ Rev.\ D} {\bf 67},  094015  (2003).
  [hep-ex/0304002].

A.~P.~Szczepaniak, M.~Swat, A.~R.~Dzierba and S.~Teige, 
  Study of the $\eta \pi$ and $\eta^\prime \pi$ spectra and interpretation of possible exotic $J^{PC} = 1^{-+}$ mesons,
 {\it Phys.\ Rev.\ Lett.}\  {\bf 91}, 092002 (2003)
  [hep-ph/0304095].

\bibitem{cleoc} G.~S.~Adams {\it et al.}  [CLEO Collaboration],
 Amplitude analyses of the decays $\chi_{c1}\to\eta \pi^+ \pi^-$ 
and $\chi_{c1}\to \eta^\prime \pi^+ \pi^-$,
  {\it Phys.\ Rev.\ D} {\bf 84}, 112009 (2011)
  [arXiv:1109.5843 [hep-ex]].
\bibitem{compass0} M.~G.~Alekseev {\it et al.} [COMPASS], Observation of a $J^{PC} = 1^{-+}$ exotic resonance in diffractive dissociation of 190-GeV/c $\pi^-$ into $\pi^- \pi^- \pi^+$, {\it Phys.\ Rev.\ Lett.}\ {\bf 104} 241803 (2010) [arXiv:1001.4654[hep-ex]].
\bibitem{compass1}
F.~Haas [COMPASS],
  Meson spectroscopy at COMPASS,
  {\it PoS EPS-HEP2009}, 081 (2009).

F.~Nerling [COMPASS], {\it PoS EPS-HEP2011}, 303 (2011) [arXiv: 1111.0259 [hep-ex]].



T.~Schl\"uter {\it et al.}  [COMPASS],
  Resonances of the systems $\pi^-\eta$ and $\pi^-\eta^\prime$ in the reactions $\pi^-p \to \pi^-\eta p$ and $\pi^-p\to \pi^-\eta^\prime p$ at COMPASS,
  {\it PoS QNP2012}, 074 (2012)
  [arXiv:1207.1076 [hep-ex]].
\bibitem{compass2} A.~Austregesilo {\it et al.}  [COMPASS],
  Partial-Wave Analysis of the Centrally Produced $\pi^+\pi^-$ System in $pp$ Reactions at COMPASS,
  {\it PoS QNP2012}, 098 (2012)
  [arXiv:1207.0949 [hep-ex]].

%\bibitem{pennington} M.~R.~Pennington,
%Understanding the baryon and meson spectra, 
%Proceedings of CIPANP 2012, St Petersburg, Florida, (May-June, 2012).



\bibitem{edwards2}
  J.~J.~Dudek and R.~G.~Edwards,
  Hybrid Baryons in QCD,
  {\it Phys.\ Rev.\ D }{\bf 85}, 054016 (2012)
  [arXiv:1201.2349 [hep-ph]].
%\bibitem{edwards3}
\bibitem{holt} R.~J.~Holt and C.~D.~Roberts, Distribution Functions of the Nucleon and Pion in the Valence Region, {\it Rev.\ Mod.\ Phys.}\ {\bf 82}, 2991 (2010) [arXiv: 1002.4666 [nucl-th]].
\bibitem{cj11} 
A.~Accardi, W.~Melnitchouk, J.~F.~Owens, M.~E.~Christy, C.~E.~Keppel, L.~Zhu and J.~G.~Morfin,
  Uncertainties in determining parton distributions at large $x$,
  {\it Phys.\ Rev.\ D} {\bf 84}, 014008 (2011)
  [arXiv:1102.3686 [hep-ph]].
\bibitem{spin1} S.~E.~Kuhn, J.-P.~Chen and E.~Leader, Spin Structure of the Nucleon --- Status and Recent Results, {\it Prog.\ Part.\ Nucl.\ Phys.}\ {\bf 63}, 1 (2009) [arXiv:0812.3535 [hep-ph]].

A.~Prokudin, Spin Structure of the proton and transverse momentum dependent distributions, {\it PoS ICHEP2010}, 167 (2010). 
\bibitem{prokudin}
  A.~Prokudin,
  Phenomenological extraction of transverse momentum dependent distributions,
  {\it AIP Conf.\ Proc.}\  {\bf 1374}, 301 (2011).

  S.~Melis, M.~Anselmino, V.~Barone, M.~Boglione, U.~D'Alesio, F.~Murgia and A.~Prokudin,
  Extraction of TMDs with global fits,
  {\it Nuovo Cim.\ C} {\bf 035N2}, 165 (2012).

\bibitem{hermes}
A.~Airapetian {\it et al.}  [HERMES],
  Observation of the Naive-T-odd Sivers Effect in Deep-Inelastic Scattering,
  {\it Phys.\ Rev.\ Lett.}\  {\bf 103}, 152002 (2009)
  [arXiv:0906.3918 [hep-ex]].
\bibitem{compass-tmd}
M.~G.~Alekseev {\it et al.}  [COMPASS],
  Measurement of the Collins and Sivers asymmetries on transversely polarised protons,
  {\it Phys.\ Lett.\ B} {\bf 692}, 240 (2010)
  [arXiv:1005.5609 [hep-ex]].
%\bibitem{halla}
%M.~Mazouz {\it et al.}, {\it Phys.\ Rev.\ Lett.}\ {\bf 99}, 242501 (2007) [arXiv:0709.0450[nucl-ex]].

\bibitem{spin2}
L.~C.~Bland, Z.-E.~Meziani, G.~Miller, M.~Vanderhaeghen, C.~Weiss and F.~Yuan, Orbital angular momentum in QCD, www.int.washington.edu/news$\underline{\;\;}$12-49w.html.

and, for example, E.~Leader, A critical assessment of angular momentum sum rules, {\it Physics Letters B} {\bf 720}, 120 (2013).

\end{thebibliography}
\end{document}